\documentclass[twocolumn,prb,aps,showpacs,a4paper,nobibnotes,footinbib]{revtex4-1}

\usepackage{hyperref}
\usepackage{epsf}
\usepackage{graphicx}
\usepackage{times}
\usepackage{multibib}

\begin{document}

\title{Stabilization of single-electron pumps by high magnetic fields}
\author{ J.~D.~Fletcher,$^{1}$ M.~Kataoka,$^{1}$ S.~P.~Giblin,$^{1}$ Sunghun Park,$^2$ H.-S. Sim,$^{2}$ P.~See,$^{1}$ T.~J.~B.~M.~Janssen,$^{1}$ J.~P.~Griffiths,$^{3}$ G.~A.~C.~Jones,$^{3}$ H.~E.~Beere,$^{3}$ and D.~A.~Ritchie$^{3}$}

\affiliation{$^{1}$ National Physical Laboratory, Hampton Road, Teddington, Middlesex TW11 0LW, United Kingdom }
\affiliation{$^{2}$ Department of Physics, Korea Advanced Institute of Science and Technology, Daejeon 305-701, Korea}
\affiliation{$^{3}$ Cavendish Laboratory, University of Cambridge, J. J. Thomson Avenue, Cambridge CB3 0HE, United Kingdom }

\date{\today}

\begin{abstract}
We demonstrate theoretically and experimentally how magnetic fields influence the single-electron tunneling dynamics in electron pumps, giving a massively enhanced quantization accuracy and providing a route to a quantum current standard based on the elementary charge. The field dependence is explained by two effects: Field-induced changes in the sensitivity of tunneling rates to the barrier potential and the suppression of nonadiabatic excitations due to a reduced sensitivity of the Fock-Darwin states to electrostatic potential.


\end{abstract}

\maketitle

\section{Introduction}

Single-electron devices proposed for quantum information technologies~\cite{elzerman2004single,hayashi2003coherent,loss1998quantum} and quantum electrical metrology~\cite{blumenthal2007gigahertz,fujiwara2008nanoampere} can be used to capture, manipulate, and release electrons through a series of gate pulses. To design such devices it is important to understand the electronic response to a rapid time-varying electrostatic potential, often in the presence of externally applied magnetic fields. The effects of magnetic confinement on electronic states~\cite{darwin1931diamagnetism,tarucha1996shell} and on electron-electron interactions~\cite{ciorga2002collapse} have been studied extensively.
However, the effect of magnetic fields on the electron dynamics in time-varying potentials is less established. Semiconductor single-electron pumps in magnetic fields are an example of a system which requires a consideration of these effects. It was found experimentally that the accuracy of the quantization current produced by these devices was strongly enhanced in magnetic field.~\cite{wright2008enhanced,kaestner2009single} More recently it has been shown how important this effect is for providing a levels of quantization accuracy (at the part per million level and below) that make these devices useful in metrological applications\cite{giblinarxiv}. The origin of this magnetic field dependence has not been explained.

We explain here how magnetic fields influence the single-electron tunneling dynamics in electron pumps and show how large magnetic fields reduce back-tunneling errors by more than five orders of magnitude.
We show that there are two distinguishable components to the field dependence of the pump accuracy. Firstly, we show through numerical calculations how magnetic fields change the back-tunneling rates in the pump; magnetic fields enhance the sensitivity of tunneling rates to the confinement barriers, which stabilizes the number of pumped electrons. Secondly, we report that the spillage of electrons through non-adiabatic processes~\cite{kataoka2011tunable}, which only appears when pumping at high frequencies, has a distinctive non-monotonic field dependence. Intriguingly, there is a also recovery of quantization accuracy at high field due to the suppression of excitations. Both effects are important in determining the ultimate current quantization accuracy in these pumps at high field, which is a crucial factor for their use in quantum metrology.~\cite{zimmerman2003electrical, giblinarxiv}

\begin{figure}
\includegraphics[width=8cm]{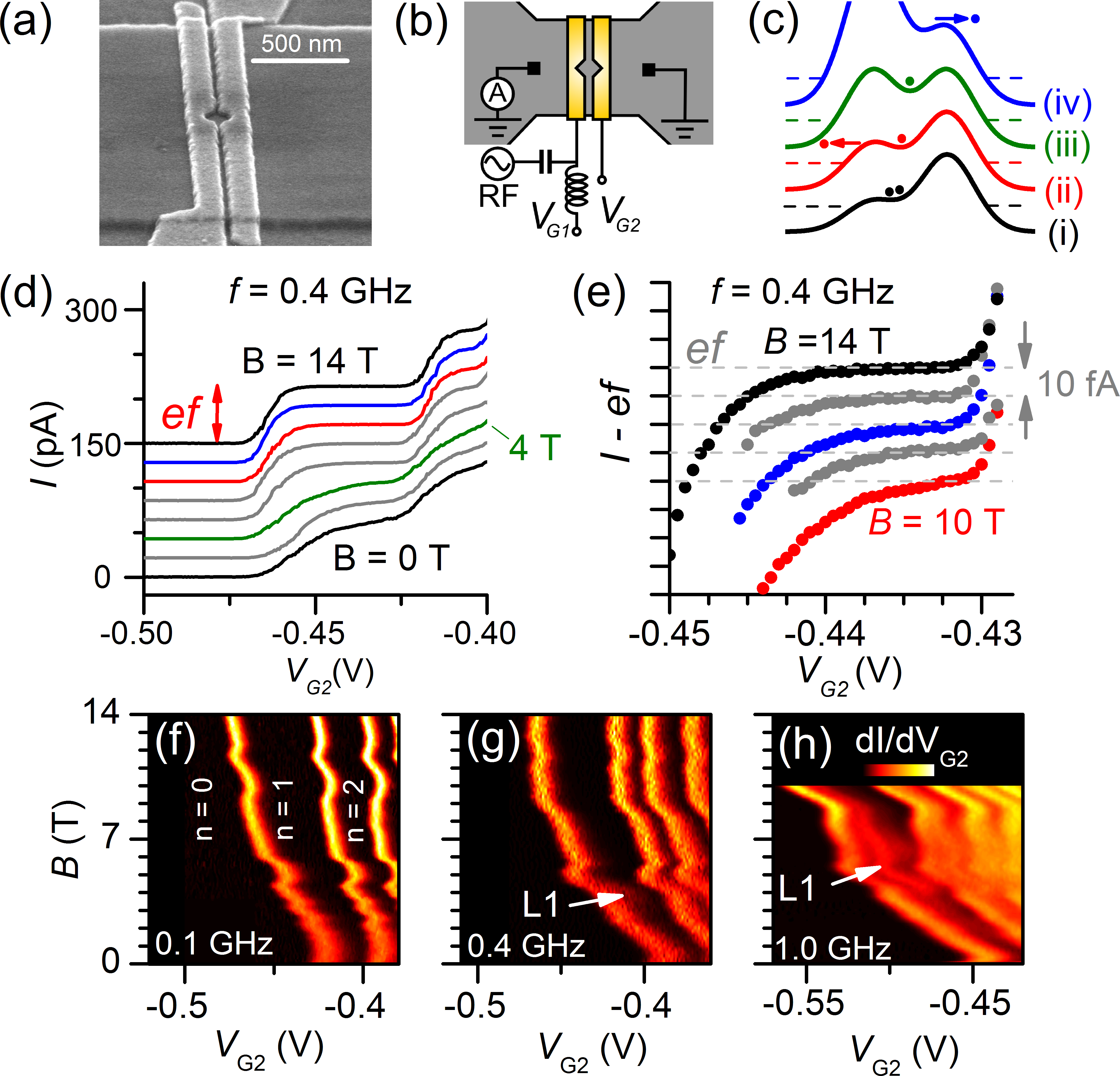}
\caption{
(a)~Scanning electron microscope image of a typical device.
(b)~Schematic electrical connections. Electrons are pumped from left to right.
(c)~Potential profile during the pumping cycle (offset vertically): i. loading, ii. back-tunneling, iii. trapping, iv. ejection.
(d)~Pump current at $f =$~0.4~GHz as a function of $V_{G2}$. Curves are offset vertically by a fixed amount as magnetic field is stepped in intervals of 2~T from 0~T to 14~T. Data for $B<$14~T have been shifted horizontally to align the $I = 1ef$ to $2ef$ transition.
(e)~High resolution scans at $f =~0.4$~GHz at $B =$ 10, 11,..14~T. Scans are offset by 10~fA. Dashed line is $ef$ for each field.
(f-h)~$dI/dV_{G2}$ on a color scale as a function of $V_{G2}$ and B for $f =$~0.1, 0.4 and 1~GHz. Structure arising from the first excited state is labeled $L_1$.}
\label{fig:sample}
\end{figure}

\section{Electron pumps in a magnetic field}

Our pumps use a dynamically formed quantum dot defined in a 2DEG AlGaAs/GaAs heterostructure~\cite{waferproperties} by two surface gates [Fig.~\ref{fig:sample}(a)]. The gates cross an etch-defined wire terminated with ohmic electrical contacts [see Fig.~\ref{fig:sample}(b)]. The potential on the \textit{entrance} gate (left) is modulated sinusoidally by $V_{\rm RF}$ around a constant value $V_{G1}$ while the \textit{exit} gate (right) is held at constant voltage $V_{G2}$.\cite{voltagevalues} Pump operation is illustrated in Fig.~\ref{fig:sample}(c)i-iv: (i)~Electrons from the source reservoir (left) are loaded into a quantum dot formed in the space between the gates. (ii)~While the dot is progressively isolated by the rising entrance barrier, some initially-trapped electrons tunnel back to the source before tunneling is eventually cut off. (iii)~The back-tunneling rate $\Gamma$ depends strongly on the number of trapped electrons $n$ due to the charging energy, leading to the same number of electrons being trapped in every cycle. (iv)~Electrons that remain trapped are forced over the exit barrier into the drain lead, producing a quantized current in an external circuit. The number of electrons pumped can be changed by adjusting the values of $V_{G1}$ and $V_{G2}$ giving current plateaux at integer multiples of $ef$, where $f$ is the operating frequency and $e$ is the elementary charge.

Figure~\ref{fig:sample}(d) shows how the pump current $I$ varies with $V_{G2}$ concentrating on the plateau at $I = nef$ with $n=1$ (one electron per cycle). Data are shown for a pump frequency $f = 0.4$~GHz in perpendicular magnetic field $B$ up to 14~T. Measurements were performed in a $^3$He cryostat with a base temperature of $\sim$~300~mK.  At higher fields the current plateaux become markedly flatter and somewhat longer, increasing the accuracy of current quantization similar to previous studies.\cite{wright2008enhanced,kaestner2009single} To allow for detailed comparison with our model we have studied these effects in several samples in detail. High-resolution measurements, using the same experimental configuration as Ref.~\onlinecite{giblinarxiv} are shown in Fig.~\ref{fig:sample}(e). Measurement for $B = 10-14$~T indicate that the pumped current gets continuously closer to the expected value of $ef$ at higher fields. This is the first time that variations with magnetic field at this level have been reported and underlines the significance of the magnetic field effect.

Figure~\ref{fig:sample}(f-h) shows $dI/dV_{G2}$ as a function of $V_{G2}$ and $B$ for $f =$~0.1, 0.4 and 1~GHz. All three data sets show the movement of plateaux boundaries in magnetic field. Some of this behavior has been linked to magnetic confinement~\cite{wright2011single}. At high magnetic fields, where we expect tunneling rates to be suppressed, it appears that a shift in $V_{G2}$ is required to recover the same current. This is consistent with the number of electrons loaded per cycle being determined by 'back-tunneling', which is sensitive to both magnetic field and the electrostatic potential. This explanation does not give any clues as to why the shape of the plateaux changes, which is also visible in this data. In fig.~\ref{fig:sample}(f) where $f = 0.1$~GHz, sharpening of the plateaux boundaries and lengthening of the plateaux are visible, massively enhancing the pump accuracy from a few percent accuracy at zero field to the part per million level.\cite{giblinarxiv} We find that at higher frequency, for instance at $f = 0.4$~GHz in \ref{fig:sample}(g), similar behavior is seen except in a certain field range (near 4~T) where step-like features appear in the $V_{G2}$ scans and broaden the plateau edge (labeled $L1$). This second effect is identified as the non-adiabatic population of excited dot states~\cite{kataoka2011tunable}. Further increasing the frequency to 1~GHz, these step features destroy plateaux flatness over a wide field range as seen in Fig. \ref{fig:sample}(h).

\begin{figure}
\includegraphics[width=8.5cm]{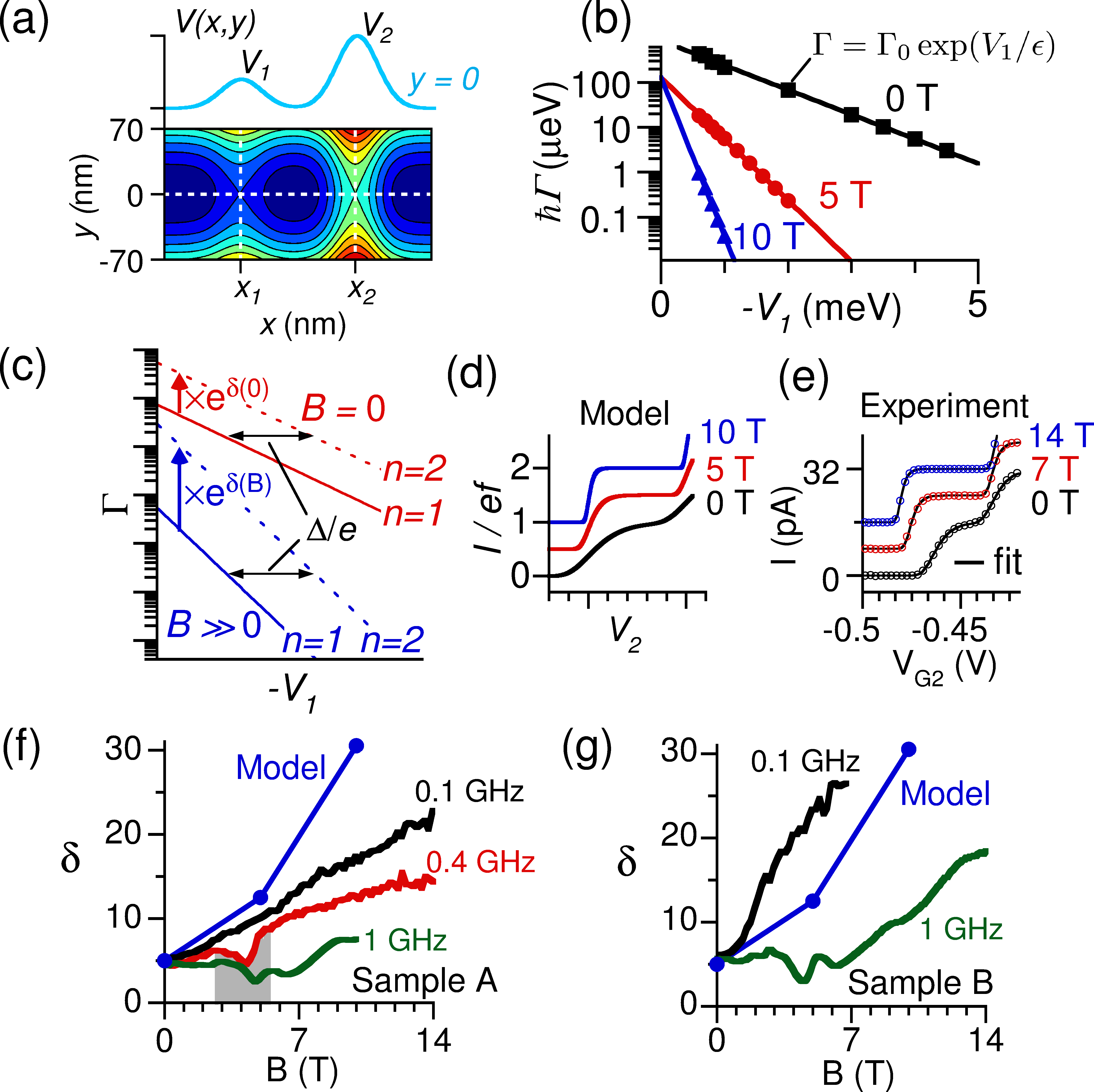}
\caption{Model calculations:
(a)~Contour plot of the model 2D potential well with a line cut along the pumping direction.
(b)~Tunneling rate $\Gamma$ from the potential well as a function of $V_1$, the entrance barrier for different values of B for a constant $V_2 = -50$~mV. Solid lines are fits to an exponential function.%
(c)~Schematic diagram indicating the relative tunneling rate for $n=1$ and $n =2$ (solid and dashed lines respectively) for states in zero and large field. $\Gamma_{2} = e^{\delta(B)}\Gamma_{1}$ and $\Gamma_{2}(eV_1) =  \Gamma_{1}(eV_1+\Delta)$. See text for definition of $\delta$.
(d)~Calculated pump current as a function of control voltage in the back-tunneling model
(e)~Fit of experimental data to determine $\delta(B)$ at 100~MHz (sample A).
(f)~Comparison of field induced changes in $\delta$ found experimentally in sample A (as in Fig.~\ref{fig:sample}) alongside model predictions (symbols). Shaded region indicates where non-adiabatic effects influence plateau flatness at $f = 0.4$~GHz (g) Similar data for sample B.}
\label{fig:model}
\end{figure}

From detailed studies of several samples we have seen that there are two field dependent contributions to the pump accuracy. At sufficiently low frequencies, where there is no evidence for excitation effects, the pump current accuracy is determined by the back-tunneling of excess electrons before the dot is isolated from the leads during the time when the entrance barrier is rising rapidly [Fig.~\ref{fig:sample}(b)~ii]. This process can be described with time-dependent tunneling rates $\Gamma_{n}(t)$ for the $n^{\rm th}$ electron out of the dot, determined by the confining geometry.\cite{KaestnerPRB08,fujiwara2008nanoampere,kashcheyevs2010universal} The disparity in tunneling rates for different numbers of electrons $\Gamma_n\ll \Gamma_{n+1}$, combined with the increasing opacity of the tunneling barrier gives a mean number of electrons captured $\hat{n} \simeq$ an integer. To understand the effect of magnetic field on the accuracy of this process a calculation of $\Gamma_n$ including the effects of magnetic confinement is required.

Below we show the results of numerical calculation for a model of the pump. We use these to show how the back-tunneling quantization process is improved in a magnetic field. We then separately consider the field dependence of non-adiabatic excitation effect, which has a different origin.

\section{Numerical calculations of tunneling rate}

Previous work has illustrated schematically the significance of the time dependent tunneling rates for the single parameter pumping using a one-dimensional model.~\cite{KaestnerPRB08, kashcheyevs2010universal} However, to include a magnetic field a two dimensional calculation is required. We have calculated the tunnel-coupling for the two-dimensional Hamiltionian $H = (i\hbar \nabla-e{\bf A})^2/2m^*-eV(x,y)$ where $m^*$ is the effective mass, ${\bf A}$ the magnetic vector potential and $V(x,y)$ the model potential
\begin{equation}-eV(x,y) = \frac{1}{2}m^*\omega_y^2y^2 - e\sum_{b=1,2} V_b \exp\left[\frac{-4(x-x_b)^2}{d^2}\right] \end{equation}
consisting of two Gaussian barriers of width $d = 60$~nm, positioned $x_2-x_1 = $120~nm apart with amplitudes $V_{1}$, $V_{2}$\cite{gatenames} and with parabolic lateral confinement $\hbar \omega_y \simeq$ 5~meV. This potential, shown in Fig.~\ref{fig:model}(a), was chosen to approximate the experimental geometry and gives an orbital energy level spacing similar to that found in our pumps.\cite{kataoka2011tunable}
The broadening of the electron energy is calculated by the lattice Green's function method.\cite{sim2001magnetic,datta1997electronic} The two-dimensional continuous system is modeled by a discrete square lattice with a tight-binding Hamiltonian~\cite{datta1997electronic}. The on-site energies of the Hamiltonian contain the position-dependent potential, while off-diagonal elements, describing the hopping between neighboring sites, include the Peierls phase factor from the magnetic field~\cite{feynman1964feynman}. The reservoir regions away from the region Fig.~\ref{fig:model}(a) are treated as semi-infinite leads.\cite{datta1997electronic}

Figure~\ref{fig:model}(b) shows the calculated back-tunneling rate $\Gamma$ as a function of $V_1$ for exit barrier height $V_2$ = -50~mV at $B =$ 0, 5, and 10~T.
This shows the expected exponential variation and a reduction in $\Gamma$ at higher fields -- the dot is increasingly decoupled from the leads by magnetic confinement. However, fitting the data to the expression $\Gamma = \Gamma_0 \exp(V_1/\epsilon)$ we can see that the \emph{sensitivity} of $\Gamma$ to the barrier height is also strongly enhanced in higher magnetic fields, with $\epsilon$  changing by a factor $\simeq 6$. We show in the next section that this effect drives a very large enhancement in quantization accuracy, but first discuss the origin of this effect. At low field the electronic wavefunction is determined solely by the electrostatic confining potential, with the penetration of the wavefunction into the confining barrier determining the sensitivity of the tunneling rate to barrier height. At high field, magnetic confinement reduces the size of the electronic wavefunction, causing the tunneling rate to drop. In our experiment this is compensated by forcing the electron closer to the barrier, but the probability density is then so concentrated that small variations in barrier height change the tunnel-coupling very rapidly.

\section{Effect on pumping accuracy}

The accuracy of the back-tunneling process is determined by the disparity in back-tunneling rates for different electron numbers. We use the above calculation for a single electron occupying the dot and use some simple approximations to deduce the effect of field on these relative tunneling rates. To estimate the tunneling rate for a state with two electrons we assume that energy of this system $E_2$ is increased by an amount $\Delta$ over the single electron case $E_1$, effectively lowering the barrier by $\Delta/e$. In this case $\Gamma_{2}$ then has the same exponential dependence on $V_1$ as $\Gamma_1$ but is shifted to higher tunneling rates by a factor~$\exp(\delta)$ where $\delta = \Delta/\epsilon e$, as shown in Fig.~\ref{fig:model}(c), where $\epsilon$ is related to the slope of the exponential behavior in Fig.~\ref{fig:model}(b). The ratio $\Gamma_{2}/\Gamma_{1}$, which determines quantization accuracy, can be enhanced either by increasing $\Delta$ (increasing the charging energy) or by increasing the sensitivity of the tunneling rate to $V_1$ (decreasing $\epsilon$). While the field dependence of the charging energy is typically observed to be very weak,~\cite{CiorgaAdditionSpectrum,tarucha1996shell,kouwenhoven2001few,Ashoori} the large changes in $\epsilon(B)$ seen in Fig.~\ref{fig:model}(b) will give large enhancements in quantization accuracy.

We show in Fig.~\ref{fig:model}(d) the calculated pump current $I(V_{2})$ using a model based on the back-tunneling process~\cite{kashcheyevs2010universal,fujiwara2008nanoampere} but including the field dependent effects found above. The functional form of $I(V_{2})$ is given by
\begin{equation}
I = ef\sum_{n=1,2} \exp\left[ -\exp \left(-\frac{\alpha (V_2-V_0)}{\epsilon}+(1-n)\delta\right)\right]
\label{eqn:cascade}
\end{equation}
where $V_0$ sets the position of the first plateau and $\delta$ sets the plateaux flatness (larger values correspond to more accurate quantization). Equation~\ref{eqn:cascade} arises from the sensitivity of $\Gamma_{1,2}$ to exit barrier height, which can be used to select the number of electrons trapped by increasing the dot energy and increasing tunneling rates.\cite{kashcheyevs2010universal} The parameter $\alpha$ is defined as the proportionality constant between exit barrier height and dot potential.

Figure~\ref{fig:model}(d) shows that, using the values of $\epsilon$ derived from our numerical calculations, Eq.~\ref{eqn:cascade} predicts a pronounced enhancement of plateaux flatness like that seen experimentally. One small difference is that while it reproduces the experimentally observed sharpening of the plateau risers, plateaux length is fixed. The change in plateaux length suggests a slight difference in the way that magnetic field enhances the sensitivity of $\Gamma_n$ to $V_2$ compared to its effect on $\Gamma(V_1)$. This would be equivalent to making $\alpha$ field dependent, which would allow the plateaux length to change with field as seen experimentally [see Fig.~\ref{fig:sample}(d)].

We fit our experimental data to Eq.~\ref{eqn:cascade} and extract an effective value of $\delta$ as a function of field to compare with our calculations.\cite{deltafit} This is shown in Fig.~\ref{fig:model}(f) and Fig.~\ref{fig:model}(g) for two samples (Sample A is the same as in Fig.~\ref{fig:sample}). At a frequency of 0.1~GHz, Eq.~\ref{eqn:cascade} fits the data well and there is no sign of any excitation effects. In both samples there is a strong monotonic increase in $\delta$. This field dependent enhancement is similar in size to that estimated in our model.

In the next section we illustrate where the scale and field dependence of this effect arises by a simple analytical estimate using the WKB model, which serves to corroborate these numerical calculations.

\section{Effect of magnetic field on tunneling rates in WKB model}

\begin{figure}
\includegraphics[width=4cm]{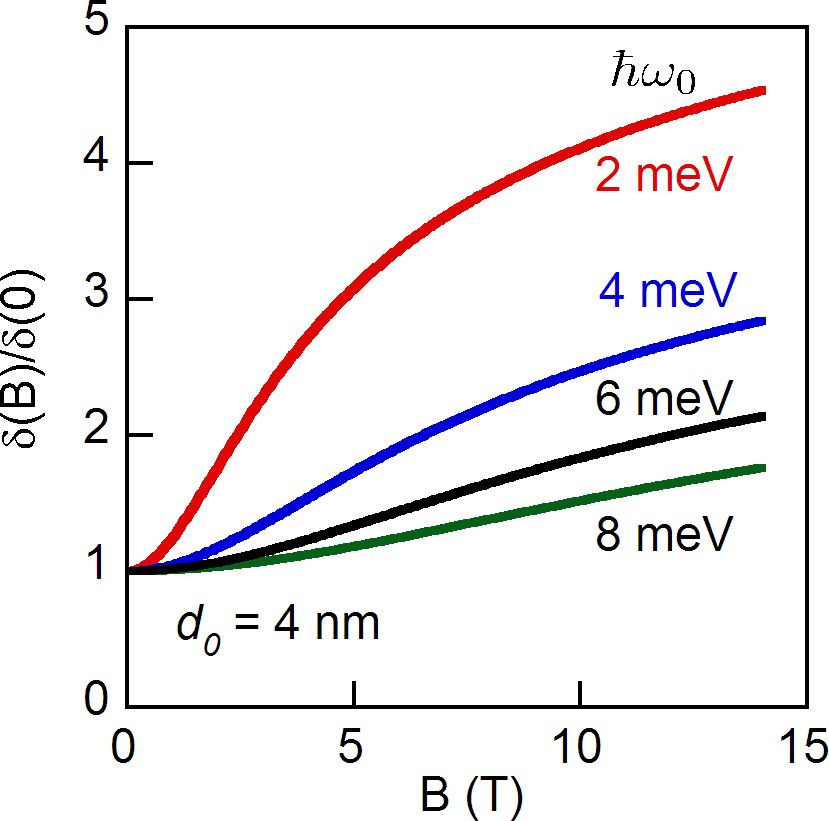}
\caption{Field dependence of $\delta(B)$ assuming that it is determined by the effective barrier thickness.}
\label{fig:de}
\end{figure}

To complement the detailed numerical calculation we can illustrate the qualitative origin of the relevant field scales and how these relate to experimental dimensions. We take a 1D WKB approximation\cite{LangerPhysRev1937} applied to the transmission coefficient for a particle tunnelling through a barrier with shape $V(x)$ and length $d$. The transmission coefficient is given by

\begin{equation}
T = \frac{\exp\left(-2 \int_{x_1}^{x_2}dx\sqrt{\frac{2m^*}{\hbar^2}\left(V(x)-E\right)}\right)}{\left[1+\frac{1}{4}\exp\left(-2 \int_{x_1}^{x_2}dx\sqrt{\frac{2m^*}{\hbar^2}\left(V(x)-E\right)}\right)\right]^2}.
\end{equation}
In the case of a constant barrier potential $V_0$ of width $d$ and considering small transmission probabilities this gives a tunneling rate
\begin{equation}
\Gamma \propto \exp \frac{-2 d \sqrt{2 m^*(V_0 - E)}}{\hbar}
\end{equation}
where an effective barrier height is defined by the quantity $V_b = (V_0 - E)$. This effective barrier can be reduced by lowering the potential barrier by an amount $\Delta V_0$, or increasing the energy of the electron by a small amount $\Delta E$, either of which will increase the tunneling rate. We then write
\begin{equation}
\Gamma \propto \exp \frac{-2 d \sqrt{2 m^*(V_0 +\Delta V_0 - E - \Delta E )}}{\hbar},
\end{equation}
expand in $\Delta V_0$ and retaining only the term that depends on $\Delta V_0$ gives us
\begin{equation}
\Gamma \propto \exp \frac{- d \sqrt{2 m^*}}{\hbar \sqrt{(V_0 - E)}}\Delta V_0
\end{equation}
showing that the sensitivity of tunneling rates to effective barrier height is determined by the thickness of the barrier. Considering for changes in energy $\Delta E$ gives an equivalent expression. We can see qualitatively that any effect which modifies this barrier thickness, in the present case a strong magnetic confinement effect, can modify the sensitivity of the tunneling rate to confinement parameters. In our model of pump operation we are normally interested in the ratio of tunneling rates $\Gamma_2/\Gamma_1 = \exp(\delta)$ from two different energy levels, corresponding to different numbers of electrons in the dot. Within the above approximations we find that
\begin{equation}
\delta = \ln\Gamma_2 - \ln\Gamma_1 = \frac{- d \sqrt{2 m^*}}{\hbar \sqrt{(V_0 - E)}}\Delta.
\end{equation}
 where the charging energy $\Delta = E_2 - E_1$ gives rise to a large difference in tunnel rates whose ratio depends on barrier geometry. To consider how the magnetic field changes the tunneling rate we introduce an effective barrier thickness $d_e(B)$ which increases with magnetic field\cite{hüttel2005direct}. This describes the fact that, under magnetic field, the spatial extent of the wavefunction is reduced, decreasing the penetration into the barrier. This will lead to an increase in the ratio of tunneling rates. In our model of pump operation this will change the plateaux quality parameter $\delta$ according to
\begin{equation}
\frac{\delta(B)}{\delta(0)} \sim \frac{d_e(B)}{d_e(0)}
\end{equation}
An estimate of the field dependence of this enhancement can be found by taking the expected changes in magnetic length  $\ell_B(B)$ from the solution to the 2D harmonic potential well in a perpendicular magnetic field\cite{kouwenhoven2001few} with $d_e = d_0 -\ell_B(B)+ \ell_B(0)$ and $\ell_B(B) = \left(m^*\sqrt{\omega_0^2+\omega_c^2/4}/\hbar^2\right)^{-1/2}$~\cite{kouwenhoven2001few} where $\omega_c = eB/m^*$ and $\hbar\omega_0$ is the electrostatic confinement strength. There is a gradual crossover from an electrostatic to magnetic dominated regime, for example for $\hbar\omega_0 = 2~$meV, $\ell_B$ shrinks from  $\sim 25$~nm at $B=0$~T to $\sim$10~nm at $B=14$~T. The result is an increase in $\delta (B)$ at high field by an amount given by the relative change in $\ell_B(B)$ compared to the zero field length $d_0$. The onset field of the effect is determined by the ratio $\hbar \omega_0/\hbar \omega_c$.

Fig.~\ref{fig:de} shows the enhancement this for a range of values of $\hbar \omega_0$ similar to that expected\cite{kataoka2011tunable} and an effective value of $d_0  = 4~nm$. Choosing a larger or smaller value of $d_0$ simply magnifies or weakens the enhancement effect, without changing the result qualitatively. This simple model reproduces the qualitative effects that are more accurately probed by the above numerical calculations.

\section{Non-adiabatic effects}

\begin{figure}
\includegraphics[width=9cm]{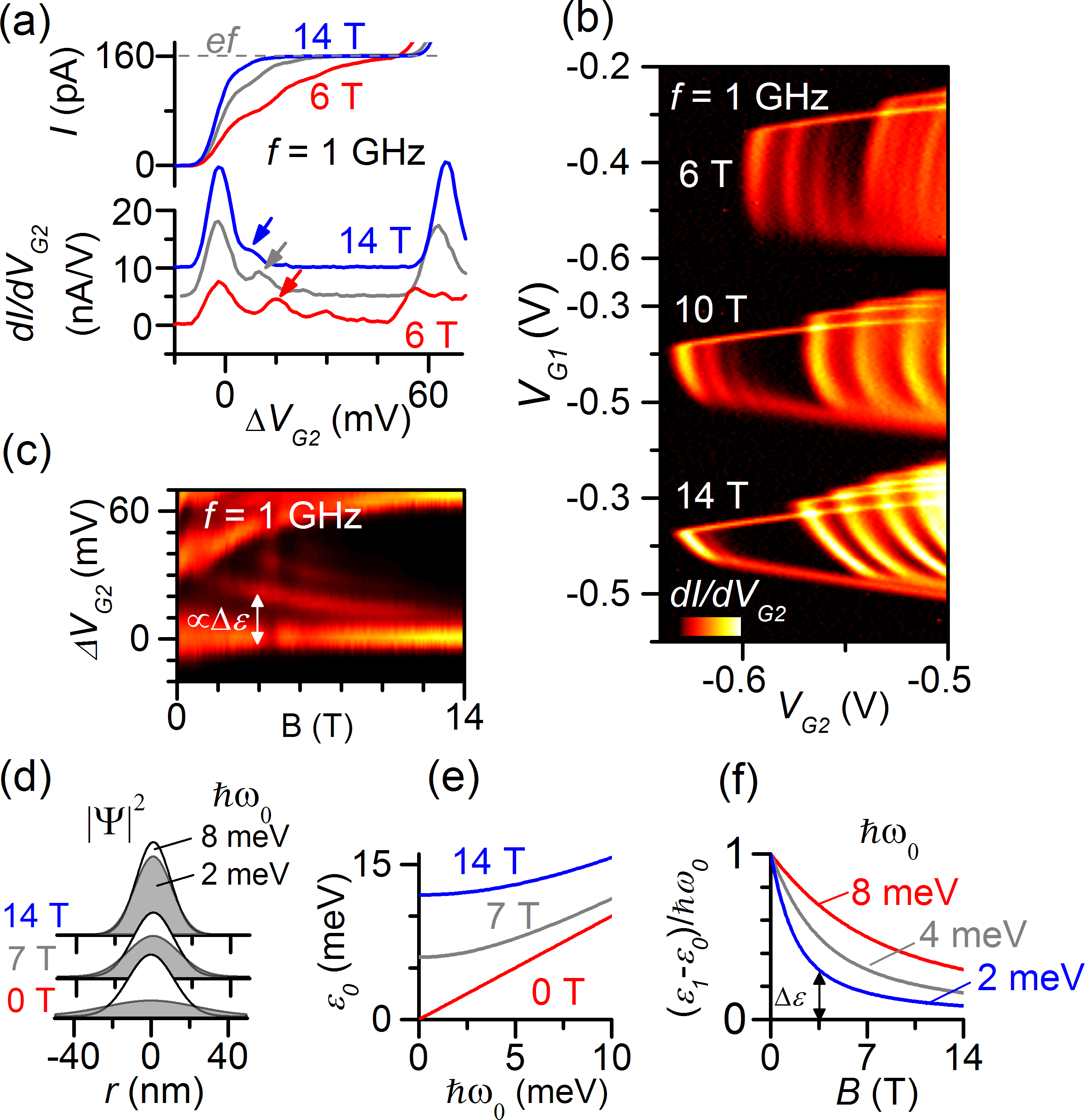}
\caption{(a) Current plateaux at 1~GHz for fields of 6, 10 and 14 T along with $dI/dV_{G2}$ (offset vertically, $V_{G1}=-0.4~V$). Voltage changes $\Delta V_{G2}$ are measured from the rising edge of the first plateau. All data in this figure is from sample B. Arrows indicate excitation features. (b) Maps of $dI/dV_{G2}$ (color scale) as a function of both $V_{G2}$ and $V_{G1}$ for 6~T, 10~T and 14~T at 1~GHz. (c) Color map of $dI/dV_{G2}(\Delta V_{G2})$ as a function of field up to 14~T. Excitation gap $\Delta \varepsilon = \varepsilon_1-\varepsilon_0$ is indicated.
(d) Fock-Darwin state probability density for different combinations of B and $\hbar\omega_0$. Open curves are for $\hbar\omega_0 = 8$~meV and filled curves are for $\hbar\omega_0 = 2$~meV.
(e) Eigenenergy of the ground state solution of the Fock-Darwin confinement potential~\cite{tarucha1996shell} for different combinations of $B$ and electrostatic confinement energy $\hbar\omega_0$.
(f) Field dependence of the first excitation gap between $(n,l) = (0,0)$ to $(n,l) = (0,1)$ states, where $n, l$ are the radial quantum number and orbital angular momentum respectively, for $\hbar\omega_0$ = 2, 4, 8~meV.}
\label{fig:excitations}
\end{figure}

Fig.~\ref{fig:model}(f) shows that at  0.4~GHz (sample A) there is also a strong increase in $\delta$ with field, although there is a difference in the maximum value of $\delta$ reached. This effect may be due to the slightly different confinement potential shape at these higher frequencies as the value of $V_{G2}$ at which the plateaux appear are different. The deterioration of pump accuracy at these higher frequencies can be overcome by the use of a specially tailored wave-form,\cite{giblinarxiv} which gives an effective five-fold increase in frequency for the same pump accuracy. A very distinctive frequency dependent effect is visible in the form of a dip in plateaux flatness around 4~T. This departure from a monotonic field dependence is associated with the onset of non-adiabatic effects. In this regime $\delta$ is strongly suppressed by excitation features, which appear as `shoulder' features in the $V_{G2}$ scans.\cite{kataoka2011tunable} Data at 1~GHz shows further suppression of $\delta$ over a wide field range, but with a recovery at high field.

In these devices rapid changes in the electrostatic confinement potential can populate excited states of the dot by nonadiabatic processes~\cite{kataoka2011tunable}, as observed in Fig.~\ref{fig:model} and in Fig.~\ref{fig:excitations}. Tunneling rates out of excited states are larger so electrons ``spill'' out of the pump and the current plateaux are eroded into a number of unquantized steps. Figure~\ref{fig:excitations}(a) shows $I(V_{G2})$ and $dI/dV_{G2}$ for $f = 1$~GHz (Sample B), high enough to induce nonadiabatic excitations of the dot. Excitation features emerge above a few Tesla~\cite{kataoka2011tunable} but as the field is increased above 12~T these features (peaks in the derivative) become significantly weaker. Figure~\ref{fig:excitations}(b) and (c) show this effect in more detail. This leads to the recovery of $\delta$ seen in Fig.~\ref{fig:model}(g).
According to Ref.\cite{kashcheyevs2010universal}, the recovery of a $\delta \simeq$ 18 is sufficient to give a quantization accuracy of $\simeq$ 3 parts in $10^{7}$. This represents an enhancement of $10^5$ over the zero field case where $\delta \sim 5$ gives only $\sim$5\% accuracy.

The observation of a limited field range where nonadiabatic effects are visible can be explained by a competition between magnetic and electrostatic effects on the electronic wavefunction. Nonadiabatic transition rates depend both on the strength and rapidity of the perturbation of the wavefunction.\cite{kataoka2011tunable} At high field $\hbar\omega_c$ (1.7~meV/T) can exceed $\hbar\omega_0 \sim$~a few meV in our system. The relative contribution of the electrostatic component is diminished and the magnetic field determines the size of the wavefunction. As a result, changes in the confinement potential during pumping have a weaker effect on the dot wavefunction and nonadiabatic transition rates are reduced. Figures~\ref{fig:excitations}(d) and \ref{fig:excitations}(e) show, for example, that the probability density $|\psi|^2$ and eigenenergy $\varepsilon_0$ of the lowest energy orbital Fock-Darwin state~\cite{kouwenhoven1997excitation} become increasingly insensitive to changes in $\omega_0$ at higher field, consistent with the disappearance of the excitation features.

At lower fields excitation features become weaker due to the significant increase in the excitation gap between ground and excited states. This field dependent gap can be seen directly in the spectrum of excitation features in Fig.~\ref{fig:excitations}(c)~\cite{kataoka2011tunable}. The excitation gap between the ground and first excited orbital energy levels $\varepsilon_1 - \varepsilon_0$ is reduced strongly with field, which may be sufficient to suppress excitations at low field. However, the magnitude of this field dependence is sensitive to $\omega_0$ [see Fig.~\ref{fig:excitations}(f)], which is time dependent in our case. Excitation processes happening earlier in the pumping cycle, when the dot confinement is smaller, would be more sensitive to magnetic fields.

In summary, we have shown that electron dynamics in single-electron tunable-barrier pumps are sensitive to magnetic fields via two mechanisms. Firstly, the increased sensitivity of tunneling rates to the confining barriers enhances the separation of back-tunneling times that is relevant for ensuring stability in the number of electrons trapped. Secondly the magnetic field plays a role in suppressing excitations of the quantum dot, which would other wise lead to unwanted spillage of trapped electrons. These effects are both important in allowing pumps to operate at high speed with error rates smaller than one part per million in high fields.~\cite{giblinarxiv}

This research is supported by the UK Department for Business, Innovation and Skills, the European Metrology Research Programme, grant no. 217257, the UK EPSRC and Korea NRF (2011-0022955).

\bibliography{fieldeffects_WOS_noURL,footnotes}

\begin{thebibliography}{29}%
\makeatletter
\providecommand \@ifxundefined [1]{%
 \@ifx{#1\undefined}
}%
\providecommand \@ifnum [1]{%
 \ifnum #1\expandafter \@firstoftwo
 \else \expandafter \@secondoftwo
 \fi
}%
\providecommand \@ifx [1]{%
 \ifx #1\expandafter \@firstoftwo
 \else \expandafter \@secondoftwo
 \fi
}%
\providecommand \natexlab [1]{#1}%
\providecommand \enquote  [1]{``#1''}%
\providecommand \bibnamefont  [1]{#1}%
\providecommand \bibfnamefont [1]{#1}%
\providecommand \citenamefont [1]{#1}%
\providecommand \href@noop [0]{\@secondoftwo}%
\providecommand \href [0]{\begingroup \@sanitize@url \@href}%
\providecommand \@href[1]{\@@startlink{#1}\@@href}%
\providecommand \@@href[1]{\endgroup#1\@@endlink}%
\providecommand \@sanitize@url [0]{\catcode `\\12\catcode `\$12\catcode
  `\&12\catcode `\#12\catcode `\^12\catcode `\_12\catcode `\%12\relax}%
\providecommand \@@startlink[1]{}%
\providecommand \@@endlink[0]{}%
\providecommand \url  [0]{\begingroup\@sanitize@url \@url }%
\providecommand \@url [1]{\endgroup\@href {#1}{\urlprefix }}%
\providecommand \urlprefix  [0]{URL }%
\providecommand \Eprint [0]{\href }%
\providecommand \doibase [0]{http://dx.doi.org/}%
\providecommand \selectlanguage [0]{\@gobble}%
\providecommand \bibinfo  [0]{\@secondoftwo}%
\providecommand \bibfield  [0]{\@secondoftwo}%
\providecommand \translation [1]{[#1]}%
\providecommand \BibitemOpen [0]{}%
\providecommand \bibitemStop [0]{}%
\providecommand \bibitemNoStop [0]{.\EOS\space}%
\providecommand \EOS [0]{\spacefactor3000\relax}%
\providecommand \BibitemShut  [1]{\csname bibitem#1\endcsname}%
\let\auto@bib@innerbib\@empty
\bibitem [{\citenamefont {Elzerman}\ \emph {et~al.}(2004)\citenamefont
  {Elzerman}, \citenamefont {Hanson}, \citenamefont {van Beveren},
  \citenamefont {Witkamp}, \citenamefont {Vandersypen},\ and\ \citenamefont
  {Kouwenhoven}}]{elzerman2004single}%
  \BibitemOpen
  \bibfield  {author} {\bibinfo {author} {\bibfnamefont {J.}~\bibnamefont
  {Elzerman}}, \bibinfo {author} {\bibfnamefont {R.}~\bibnamefont {Hanson}},
  \bibinfo {author} {\bibfnamefont {L.}~\bibnamefont {van Beveren}}, \bibinfo
  {author} {\bibfnamefont {B.}~\bibnamefont {Witkamp}}, \bibinfo {author}
  {\bibfnamefont {L.}~\bibnamefont {Vandersypen}}, \ and\ \bibinfo {author}
  {\bibfnamefont {L.}~\bibnamefont {Kouwenhoven}},\ }\href {\doibase
  10.1038/nature02693} {\bibfield  {journal} {\bibinfo  {journal} {Nature}\
  }\textbf {\bibinfo {volume} {430}},\ \bibinfo {pages} {431} (\bibinfo {year}
  {2004})}\BibitemShut {NoStop}%
\bibitem [{\citenamefont {Hayashi}\ \emph {et~al.}(2003)\citenamefont
  {Hayashi}, \citenamefont {Fujisawa}, \citenamefont {Cheong}, \citenamefont
  {Jeong},\ and\ \citenamefont {Hirayama}}]{hayashi2003coherent}%
  \BibitemOpen
  \bibfield  {author} {\bibinfo {author} {\bibfnamefont {T.}~\bibnamefont
  {Hayashi}}, \bibinfo {author} {\bibfnamefont {T.}~\bibnamefont {Fujisawa}},
  \bibinfo {author} {\bibfnamefont {H.}~\bibnamefont {Cheong}}, \bibinfo
  {author} {\bibfnamefont {Y.}~\bibnamefont {Jeong}}, \ and\ \bibinfo {author}
  {\bibfnamefont {Y.}~\bibnamefont {Hirayama}},\ }\href {\doibase
  10.1103/PhysRevLett.91.226804} {\bibfield  {journal} {\bibinfo  {journal}
  {Phys. Rev. Lett.}\ }\textbf {\bibinfo {volume} {91}},\ \bibinfo {pages}
  {226804} (\bibinfo {year} {2003})}\BibitemShut {NoStop}%
\bibitem [{\citenamefont {Loss}\ and\ \citenamefont
  {DiVincenzo}(1998)}]{loss1998quantum}%
  \BibitemOpen
  \bibfield  {author} {\bibinfo {author} {\bibfnamefont {D.}~\bibnamefont
  {Loss}}\ and\ \bibinfo {author} {\bibfnamefont {D.}~\bibnamefont
  {DiVincenzo}},\ }\href {\doibase 10.1103/PhysRevA.57.120} {\bibfield
  {journal} {\bibinfo  {journal} {Phys. Rev. A}\ }\textbf {\bibinfo {volume}
  {57}},\ \bibinfo {pages} {120} (\bibinfo {year} {1998})}\BibitemShut
  {NoStop}%
\bibitem [{\citenamefont {Blumenthal}\ \emph {et~al.}(2007)\citenamefont
  {Blumenthal}, \citenamefont {Kaestner}, \citenamefont {Li}, \citenamefont
  {Giblin}, \citenamefont {Janssen}, \citenamefont {Pepper}, \citenamefont
  {Anderson}, \citenamefont {Jones},\ and\ \citenamefont
  {Ritchie}}]{blumenthal2007gigahertz}%
  \BibitemOpen
  \bibfield  {author} {\bibinfo {author} {\bibfnamefont {M.~D.}\ \bibnamefont
  {Blumenthal}}, \bibinfo {author} {\bibfnamefont {B.}~\bibnamefont
  {Kaestner}}, \bibinfo {author} {\bibfnamefont {L.}~\bibnamefont {Li}},
  \bibinfo {author} {\bibfnamefont {S.}~\bibnamefont {Giblin}}, \bibinfo
  {author} {\bibfnamefont {T.~J. B.~M.}\ \bibnamefont {Janssen}}, \bibinfo
  {author} {\bibfnamefont {M.}~\bibnamefont {Pepper}}, \bibinfo {author}
  {\bibfnamefont {D.}~\bibnamefont {Anderson}}, \bibinfo {author}
  {\bibfnamefont {G.}~\bibnamefont {Jones}}, \ and\ \bibinfo {author}
  {\bibfnamefont {D.~A.}\ \bibnamefont {Ritchie}},\ }\href {\doibase
  10.1038/nphys582} {\bibfield  {journal} {\bibinfo  {journal} {Nat. Phys.}\
  }\textbf {\bibinfo {volume} {3}},\ \bibinfo {pages} {343} (\bibinfo {year}
  {2007})}\BibitemShut {NoStop}%
\bibitem [{\citenamefont {Fujiwara}\ \emph {et~al.}(2008)\citenamefont
  {Fujiwara}, \citenamefont {Nishiguchi},\ and\ \citenamefont
  {Ono}}]{fujiwara2008nanoampere}%
  \BibitemOpen
  \bibfield  {author} {\bibinfo {author} {\bibfnamefont {A.}~\bibnamefont
  {Fujiwara}}, \bibinfo {author} {\bibfnamefont {K.}~\bibnamefont
  {Nishiguchi}}, \ and\ \bibinfo {author} {\bibfnamefont {Y.}~\bibnamefont
  {Ono}},\ }\href {\doibase 10.1063/1.2837544} {\bibfield  {journal} {\bibinfo
  {journal} {Appl. Phys. Lett.}\ }\textbf {\bibinfo {volume} {92}},\ \bibinfo
  {pages} {042102} (\bibinfo {year} {2008})}\BibitemShut {NoStop}%
\bibitem [{\citenamefont {Darwin}(1931)}]{darwin1931diamagnetism}%
  \BibitemOpen
  \bibfield  {author} {\bibinfo {author} {\bibfnamefont {C.}~\bibnamefont
  {Darwin}},\ }in\ \href {\doibase 10.1017/S0305004100009373} {\emph {\bibinfo
  {booktitle} {Mathematical Proceedings of the Cambridge Philosophical
  Society}}},\ Vol.~\bibinfo {volume} {27}\ (\bibinfo {organization} {Cambridge
  Univ Press},\ \bibinfo {year} {1931})\ p.~\bibinfo {pages} {86}\BibitemShut
  {NoStop}%
\bibitem [{\citenamefont {Tarucha}\ \emph {et~al.}(1996)\citenamefont
  {Tarucha}, \citenamefont {Austing}, \citenamefont {Honda}, \citenamefont
  {vanderHage},\ and\ \citenamefont {Kouwenhoven}}]{tarucha1996shell}%
  \BibitemOpen
  \bibfield  {author} {\bibinfo {author} {\bibfnamefont {S.}~\bibnamefont
  {Tarucha}}, \bibinfo {author} {\bibfnamefont {D.}~\bibnamefont {Austing}},
  \bibinfo {author} {\bibfnamefont {T.}~\bibnamefont {Honda}}, \bibinfo
  {author} {\bibfnamefont {R.}~\bibnamefont {vanderHage}}, \ and\ \bibinfo
  {author} {\bibfnamefont {L.}~\bibnamefont {Kouwenhoven}},\ }\href {\doibase
  10.1103/PhysRevLett.77.3613} {\bibfield  {journal} {\bibinfo  {journal}
  {Phys. Rev. Lett.}\ }\textbf {\bibinfo {volume} {77}},\ \bibinfo {pages}
  {3613} (\bibinfo {year} {1996})}\BibitemShut {NoStop}%
\bibitem [{\citenamefont {Ciorga}\ \emph {et~al.}(2002)\citenamefont {Ciorga},
  \citenamefont {Wensauer}, \citenamefont {Pioro-Ladriere}, \citenamefont
  {Korkusinski}, \citenamefont {Kyriakidis}, \citenamefont {Sachrajda},\ and\
  \citenamefont {Hawrylak}}]{ciorga2002collapse}%
  \BibitemOpen
  \bibfield  {author} {\bibinfo {author} {\bibfnamefont {M.}~\bibnamefont
  {Ciorga}}, \bibinfo {author} {\bibfnamefont {A.}~\bibnamefont {Wensauer}},
  \bibinfo {author} {\bibfnamefont {M.}~\bibnamefont {Pioro-Ladriere}},
  \bibinfo {author} {\bibfnamefont {M.}~\bibnamefont {Korkusinski}}, \bibinfo
  {author} {\bibfnamefont {J.}~\bibnamefont {Kyriakidis}}, \bibinfo {author}
  {\bibfnamefont {A.}~\bibnamefont {Sachrajda}}, \ and\ \bibinfo {author}
  {\bibfnamefont {P.}~\bibnamefont {Hawrylak}},\ }\href {\doibase
  10.1103/PhysRevLett.88.256804} {\bibfield  {journal} {\bibinfo  {journal}
  {Phys. Rev. Lett.}\ }\textbf {\bibinfo {volume} {88}},\ \bibinfo {pages}
  {256804} (\bibinfo {year} {2002})}\BibitemShut {NoStop}%
\bibitem [{\citenamefont {Wright}\ \emph {et~al.}(2008)\citenamefont {Wright},
  \citenamefont {Blumenthal}, \citenamefont {Gumbs}, \citenamefont {Thorn},
  \citenamefont {Pepper}, \citenamefont {Janssen}, \citenamefont {Holmes},
  \citenamefont {Anderson}, \citenamefont {Jones}, \citenamefont {Nicoll},\
  and\ \citenamefont {Ritchie}}]{wright2008enhanced}%
  \BibitemOpen
  \bibfield  {author} {\bibinfo {author} {\bibfnamefont {S.~J.}\ \bibnamefont
  {Wright}}, \bibinfo {author} {\bibfnamefont {M.~D.}\ \bibnamefont
  {Blumenthal}}, \bibinfo {author} {\bibfnamefont {G.}~\bibnamefont {Gumbs}},
  \bibinfo {author} {\bibfnamefont {A.~L.}\ \bibnamefont {Thorn}}, \bibinfo
  {author} {\bibfnamefont {M.}~\bibnamefont {Pepper}}, \bibinfo {author}
  {\bibfnamefont {T.~J. B.~M.}\ \bibnamefont {Janssen}}, \bibinfo {author}
  {\bibfnamefont {S.~N.}\ \bibnamefont {Holmes}}, \bibinfo {author}
  {\bibfnamefont {D.}~\bibnamefont {Anderson}}, \bibinfo {author}
  {\bibfnamefont {G.~A.~C.}\ \bibnamefont {Jones}}, \bibinfo {author}
  {\bibfnamefont {C.~A.}\ \bibnamefont {Nicoll}}, \ and\ \bibinfo {author}
  {\bibfnamefont {D.~A.}\ \bibnamefont {Ritchie}},\ }\href {\doibase
  10.1103/PhysRevB.78.233311} {\bibfield  {journal} {\bibinfo  {journal} {Phys.
  Rev. B}\ }\textbf {\bibinfo {volume} {78}},\ \bibinfo {pages} {233311}
  (\bibinfo {year} {2008})}\BibitemShut {NoStop}%
\bibitem [{\citenamefont {Kaestner}\ \emph {et~al.}(2009)\citenamefont
  {Kaestner}, \citenamefont {Leicht}, \citenamefont {Kashcheyevs},
  \citenamefont {Pierz}, \citenamefont {Siegner},\ and\ \citenamefont
  {Schumacher}}]{kaestner2009single}%
  \BibitemOpen
  \bibfield  {author} {\bibinfo {author} {\bibfnamefont {B.}~\bibnamefont
  {Kaestner}}, \bibinfo {author} {\bibfnamefont {C.}~\bibnamefont {Leicht}},
  \bibinfo {author} {\bibfnamefont {V.}~\bibnamefont {Kashcheyevs}}, \bibinfo
  {author} {\bibfnamefont {K.}~\bibnamefont {Pierz}}, \bibinfo {author}
  {\bibfnamefont {U.}~\bibnamefont {Siegner}}, \ and\ \bibinfo {author}
  {\bibfnamefont {H.~W.}\ \bibnamefont {Schumacher}},\ }\href {\doibase
  10.1063/1.3063128} {\bibfield  {journal} {\bibinfo  {journal} {Appl. Phys.
  Lett.}\ }\textbf {\bibinfo {volume} {94}},\ \bibinfo {pages} {012106}
  (\bibinfo {year} {2009})}\BibitemShut {NoStop}%
\bibitem [{\citenamefont {Giblin}\ \emph {et~al.}(2012)\citenamefont {Giblin},
  \citenamefont {Kataoka}, \citenamefont {Fletcher}, \citenamefont {See},
  \citenamefont {Janssen}, \citenamefont {Griffiths}, \citenamefont {Jones},
  \citenamefont {Farrer},\ and\ \citenamefont {Ritchie}}]{giblinarxiv}%
  \BibitemOpen
  \bibfield  {author} {\bibinfo {author} {\bibfnamefont {S.}~\bibnamefont
  {Giblin}}, \bibinfo {author} {\bibfnamefont {M.}~\bibnamefont {Kataoka}},
  \bibinfo {author} {\bibfnamefont {J.}~\bibnamefont {Fletcher}}, \bibinfo
  {author} {\bibfnamefont {P.}~\bibnamefont {See}}, \bibinfo {author}
  {\bibfnamefont {T.}~\bibnamefont {Janssen}}, \bibinfo {author} {\bibfnamefont
  {J.}~\bibnamefont {Griffiths}}, \bibinfo {author} {\bibfnamefont
  {G.}~\bibnamefont {Jones}}, \bibinfo {author} {\bibfnamefont
  {I.}~\bibnamefont {Farrer}}, \ and\ \bibinfo {author} {\bibfnamefont
  {D.}~\bibnamefont {Ritchie}},\ }\href {http://dx.doi.org/10.1038/ncomms1935}
  {\bibfield  {journal} {\bibinfo  {journal} {Nat Commun}\ }\textbf {\bibinfo
  {volume} {3}},\ \bibinfo {pages} {930} (\bibinfo {year} {2012})}\BibitemShut
  {NoStop}%
\bibitem [{\citenamefont {Kataoka}\ \emph {et~al.}(2011)\citenamefont
  {Kataoka}, \citenamefont {Fletcher}, \citenamefont {See}, \citenamefont
  {Giblin}, \citenamefont {Janssen}, \citenamefont {Griffiths}, \citenamefont
  {Jones}, \citenamefont {Farrer},\ and\ \citenamefont
  {Ritchie}}]{kataoka2011tunable}%
  \BibitemOpen
  \bibfield  {author} {\bibinfo {author} {\bibfnamefont {M.}~\bibnamefont
  {Kataoka}}, \bibinfo {author} {\bibfnamefont {J.~D.}\ \bibnamefont
  {Fletcher}}, \bibinfo {author} {\bibfnamefont {P.}~\bibnamefont {See}},
  \bibinfo {author} {\bibfnamefont {S.~P.}\ \bibnamefont {Giblin}}, \bibinfo
  {author} {\bibfnamefont {T.~J. B.~M.}\ \bibnamefont {Janssen}}, \bibinfo
  {author} {\bibfnamefont {J.~P.}\ \bibnamefont {Griffiths}}, \bibinfo {author}
  {\bibfnamefont {G.~A.~C.}\ \bibnamefont {Jones}}, \bibinfo {author}
  {\bibfnamefont {I.}~\bibnamefont {Farrer}}, \ and\ \bibinfo {author}
  {\bibfnamefont {D.~A.}\ \bibnamefont {Ritchie}},\ }\href {\doibase
  10.1103/PhysRevLett.106.126801} {\bibfield  {journal} {\bibinfo  {journal}
  {Phys. Rev. Lett.}\ }\textbf {\bibinfo {volume} {106}},\ \bibinfo {pages}
  {126801} (\bibinfo {year} {2011})}\BibitemShut {NoStop}%
\bibitem [{\citenamefont {Zimmerman}\ and\ \citenamefont
  {Keller}(2003)}]{zimmerman2003electrical}%
  \BibitemOpen
  \bibfield  {author} {\bibinfo {author} {\bibfnamefont {N.~M.}\ \bibnamefont
  {Zimmerman}}\ and\ \bibinfo {author} {\bibfnamefont {M.~W.}\ \bibnamefont
  {Keller}},\ }\href {\doibase 10.1088/0957-0233/14/8/307} {\bibfield
  {journal} {\bibinfo  {journal} {Measurement Science and Technology}\ }\textbf
  {\bibinfo {volume} {14}},\ \bibinfo {pages} {1237} (\bibinfo {year}
  {2003})}\BibitemShut {NoStop}%
\bibitem [{waf()}]{waferproperties}%
  \BibitemOpen
  \href@noop {} {}\bibinfo {note} {Carrier density $2.7\times 10^{15}$ m$^2$ at
  1.5 K, mobility is 70~m$^2$/Vs}\BibitemShut {NoStop}%
\bibitem [{vol()}]{voltagevalues}%
  \BibitemOpen
  \href@noop {} {}\bibinfo {note} {Peak to peak $V_{\rm RF} \simeq
  V_{G1},V_{G2} \simeq 0.4~$V}\BibitemShut {NoStop}%
\bibitem [{\citenamefont {Wright}\ \emph {et~al.}(2011)\citenamefont {Wright},
  \citenamefont {Thorn}, \citenamefont {Blumenthal}, \citenamefont {Giblin},
  \citenamefont {Pepper}, \citenamefont {Janssen}, \citenamefont {Kataoka},
  \citenamefont {Fletcher}, \citenamefont {Jones}, \citenamefont {Nicoll},
  \citenamefont {Gumbs},\ and\ \citenamefont {Ritchie}}]{wright2011single}%
  \BibitemOpen
  \bibfield  {author} {\bibinfo {author} {\bibfnamefont {S.~J.}\ \bibnamefont
  {Wright}}, \bibinfo {author} {\bibfnamefont {A.~L.}\ \bibnamefont {Thorn}},
  \bibinfo {author} {\bibfnamefont {M.~D.}\ \bibnamefont {Blumenthal}},
  \bibinfo {author} {\bibfnamefont {S.~P.}\ \bibnamefont {Giblin}}, \bibinfo
  {author} {\bibfnamefont {M.}~\bibnamefont {Pepper}}, \bibinfo {author}
  {\bibfnamefont {T.~J. B.~M.}\ \bibnamefont {Janssen}}, \bibinfo {author}
  {\bibfnamefont {M.}~\bibnamefont {Kataoka}}, \bibinfo {author} {\bibfnamefont
  {J.~D.}\ \bibnamefont {Fletcher}}, \bibinfo {author} {\bibfnamefont
  {G.~A.~C.}\ \bibnamefont {Jones}}, \bibinfo {author} {\bibfnamefont {C.~A.}\
  \bibnamefont {Nicoll}}, \bibinfo {author} {\bibfnamefont {G.}~\bibnamefont
  {Gumbs}}, \ and\ \bibinfo {author} {\bibfnamefont {D.~A.}\ \bibnamefont
  {Ritchie}},\ }\href {\doibase 10.1063/1.3578685} {\bibfield  {journal}
  {\bibinfo  {journal} {J. Appl. Phys.}\ }\textbf {\bibinfo {volume} {109}},\
  \bibinfo {pages} {102422} (\bibinfo {year} {2011})}\BibitemShut {NoStop}%
\bibitem [{\citenamefont {Kaestner}\ \emph {et~al.}(2008)\citenamefont
  {Kaestner}, \citenamefont {Kashcheyevs}, \citenamefont {Amakawa},
  \citenamefont {Blumenthal}, \citenamefont {Li}, \citenamefont {Janssen},
  \citenamefont {Hein}, \citenamefont {Pierz}, \citenamefont {Weimann},
  \citenamefont {Siegner},\ and\ \citenamefont {Schumacher}}]{KaestnerPRB08}%
  \BibitemOpen
  \bibfield  {author} {\bibinfo {author} {\bibfnamefont {B.}~\bibnamefont
  {Kaestner}}, \bibinfo {author} {\bibfnamefont {V.}~\bibnamefont
  {Kashcheyevs}}, \bibinfo {author} {\bibfnamefont {S.}~\bibnamefont
  {Amakawa}}, \bibinfo {author} {\bibfnamefont {M.~D.}\ \bibnamefont
  {Blumenthal}}, \bibinfo {author} {\bibfnamefont {L.}~\bibnamefont {Li}},
  \bibinfo {author} {\bibfnamefont {T.~J. B.~M.}\ \bibnamefont {Janssen}},
  \bibinfo {author} {\bibfnamefont {G.}~\bibnamefont {Hein}}, \bibinfo {author}
  {\bibfnamefont {K.}~\bibnamefont {Pierz}}, \bibinfo {author} {\bibfnamefont
  {T.}~\bibnamefont {Weimann}}, \bibinfo {author} {\bibfnamefont
  {U.}~\bibnamefont {Siegner}}, \ and\ \bibinfo {author} {\bibfnamefont
  {H.~W.}\ \bibnamefont {Schumacher}},\ }\href {\doibase
  10.1103/PhysRevB.77.153301} {\bibfield  {journal} {\bibinfo  {journal} {Phys.
  Rev. B}\ }\textbf {\bibinfo {volume} {77}},\ \bibinfo {pages} {153301}
  (\bibinfo {year} {2008})}\BibitemShut {NoStop}%
\bibitem [{\citenamefont {Kashcheyevs}\ and\ \citenamefont
  {Kaestner}(2010)}]{kashcheyevs2010universal}%
  \BibitemOpen
  \bibfield  {author} {\bibinfo {author} {\bibfnamefont {V.}~\bibnamefont
  {Kashcheyevs}}\ and\ \bibinfo {author} {\bibfnamefont {B.}~\bibnamefont
  {Kaestner}},\ }\href {\doibase 10.1103/PhysRevLett.104.186805} {\bibfield
  {journal} {\bibinfo  {journal} {Phys. Rev. Lett.}\ }\textbf {\bibinfo
  {volume} {104}},\ \bibinfo {pages} {186805} (\bibinfo {year}
  {2010})}\BibitemShut {NoStop}%
\bibitem [{gat()}]{gatenames}%
  \BibitemOpen
  \href@noop {} {}\bibinfo {note} {We distinguish the numerical values
  $V_1,V_2$ of potential within the 2DEG from the experimental voltages
  $V_{G1,G2}$ applied to surface gates.}\BibitemShut {Stop}%
\bibitem [{\citenamefont {Sim}\ \emph {et~al.}(2001)\citenamefont {Sim},
  \citenamefont {Ihm}, \citenamefont {Kim},\ and\ \citenamefont
  {Chang}}]{sim2001magnetic}%
  \BibitemOpen
  \bibfield  {author} {\bibinfo {author} {\bibfnamefont {H.-S.}\ \bibnamefont
  {Sim}}, \bibinfo {author} {\bibfnamefont {G.}~\bibnamefont {Ihm}}, \bibinfo
  {author} {\bibfnamefont {N.}~\bibnamefont {Kim}}, \ and\ \bibinfo {author}
  {\bibfnamefont {K.}~\bibnamefont {Chang}},\ }\href {\doibase
  10.1103/PhysRevLett.87.146601} {\bibfield  {journal} {\bibinfo  {journal}
  {Phys. Rev. Lett.}\ }\textbf {\bibinfo {volume} {87}},\ \bibinfo {pages}
  {146601} (\bibinfo {year} {2001})}\BibitemShut {NoStop}%
\bibitem [{\citenamefont {Datta}(1997)}]{datta1997electronic}%
  \BibitemOpen
  \bibfield  {author} {\bibinfo {author} {\bibfnamefont {S.}~\bibnamefont
  {Datta}},\ }\href {\doibase 10.2277/ 0521599431} {\emph {\bibinfo {title}
  {Electronic transport in mesoscopic systems}}}\ (\bibinfo  {publisher}
  {Cambridge Univ Press},\ \bibinfo {year} {1997})\BibitemShut {NoStop}%
\bibitem [{\citenamefont {Feynman}\ \emph {et~al.}(1964)\citenamefont
  {Feynman}, \citenamefont {Leighton}, \citenamefont {Sands} \emph
  {et~al.}}]{feynman1964feynman}%
  \BibitemOpen
  \bibfield  {author} {\bibinfo {author} {\bibfnamefont {R.}~\bibnamefont
  {Feynman}}, \bibinfo {author} {\bibfnamefont {R.}~\bibnamefont {Leighton}},
  \bibinfo {author} {\bibfnamefont {M.}~\bibnamefont {Sands}},  \emph
  {et~al.},\ }\href {http://dx.doi.org/} {\emph {\bibinfo {title} {The Feynman
  lectures on physics III}}}\ (\bibinfo  {publisher} {Addison-Wesley Reading,
  MA},\ \bibinfo {year} {1964})\BibitemShut {NoStop}%
\bibitem [{\citenamefont {Ciorga}\ \emph {et~al.}(2000)\citenamefont {Ciorga},
  \citenamefont {Sachrajda}, \citenamefont {Hawrylak}, \citenamefont {Gould},
  \citenamefont {Zawadzki}, \citenamefont {Jullian}, \citenamefont {Feng},\
  and\ \citenamefont {Wasilewski}}]{CiorgaAdditionSpectrum}%
  \BibitemOpen
  \bibfield  {author} {\bibinfo {author} {\bibfnamefont {M.}~\bibnamefont
  {Ciorga}}, \bibinfo {author} {\bibfnamefont {A.~S.}\ \bibnamefont
  {Sachrajda}}, \bibinfo {author} {\bibfnamefont {P.}~\bibnamefont {Hawrylak}},
  \bibinfo {author} {\bibfnamefont {C.}~\bibnamefont {Gould}}, \bibinfo
  {author} {\bibfnamefont {P.}~\bibnamefont {Zawadzki}}, \bibinfo {author}
  {\bibfnamefont {S.}~\bibnamefont {Jullian}}, \bibinfo {author} {\bibfnamefont
  {Y.}~\bibnamefont {Feng}}, \ and\ \bibinfo {author} {\bibfnamefont
  {Z.}~\bibnamefont {Wasilewski}},\ }\href {\doibase
  10.1103/PhysRevB.61.R16315} {\bibfield  {journal} {\bibinfo  {journal} {Phys.
  Rev. B}\ }\textbf {\bibinfo {volume} {61}},\ \bibinfo {pages} {R16315}
  (\bibinfo {year} {2000})}\BibitemShut {NoStop}%
\bibitem [{\citenamefont {Kouwenhoven}\ \emph {et~al.}(2001)\citenamefont
  {Kouwenhoven}, \citenamefont {Austing},\ and\ \citenamefont
  {Tarucha}}]{kouwenhoven2001few}%
  \BibitemOpen
  \bibfield  {author} {\bibinfo {author} {\bibfnamefont {L.}~\bibnamefont
  {Kouwenhoven}}, \bibinfo {author} {\bibfnamefont {D.}~\bibnamefont
  {Austing}}, \ and\ \bibinfo {author} {\bibfnamefont {S.}~\bibnamefont
  {Tarucha}},\ }\href {\doibase 10.1088/0034-4885/64/6/201} {\bibfield
  {journal} {\bibinfo  {journal} {Reports on Progress in Physics}\ }\textbf
  {\bibinfo {volume} {64}},\ \bibinfo {pages} {701} (\bibinfo {year}
  {2001})}\BibitemShut {NoStop}%
\bibitem [{\citenamefont {Ashoori}\ \emph {et~al.}(1993)\citenamefont
  {Ashoori}, \citenamefont {Stormer}, \citenamefont {Weiner}, \citenamefont
  {Pfeiffer}, \citenamefont {Baldwin},\ and\ \citenamefont {West}}]{Ashoori}%
  \BibitemOpen
  \bibfield  {author} {\bibinfo {author} {\bibfnamefont {R.~C.}\ \bibnamefont
  {Ashoori}}, \bibinfo {author} {\bibfnamefont {H.~L.}\ \bibnamefont
  {Stormer}}, \bibinfo {author} {\bibfnamefont {J.~S.}\ \bibnamefont {Weiner}},
  \bibinfo {author} {\bibfnamefont {L.~N.}\ \bibnamefont {Pfeiffer}}, \bibinfo
  {author} {\bibfnamefont {K.~W.}\ \bibnamefont {Baldwin}}, \ and\ \bibinfo
  {author} {\bibfnamefont {K.~W.}\ \bibnamefont {West}},\ }\href {\doibase
  10.1103/PhysRevLett.71.613} {\bibfield  {journal} {\bibinfo  {journal} {Phys.
  Rev. Lett.}\ }\textbf {\bibinfo {volume} {71}},\ \bibinfo {pages} {613}
  (\bibinfo {year} {1993})}\BibitemShut {NoStop}%
\bibitem [{del()}]{deltafit}%
  \BibitemOpen
  \href@noop {} {}\bibinfo {note} {The experimental scaling of $V_{G2}$ which
  is $(\alpha/\epsilon)$ is also treated as a free parameter. In the
  calculations the zero field value of $\delta$ is set at the experimental
  value.}\BibitemShut {Stop}%
\bibitem [{\citenamefont {Langer}(1937)}]{LangerPhysRev1937}%
  \BibitemOpen
  \bibfield  {author} {\bibinfo {author} {\bibfnamefont {R.~E.}\ \bibnamefont
  {Langer}},\ }\href {\doibase 10.1103/PhysRev.51.669} {\bibfield  {journal}
  {\bibinfo  {journal} {Phys. Rev.}\ }\textbf {\bibinfo {volume} {51}},\
  \bibinfo {pages} {669} (\bibinfo {year} {1937})}\BibitemShut {NoStop}%
\bibitem [{\citenamefont {H{\"u}ttel}\ \emph {et~al.}(2005)\citenamefont
  {H{\"u}ttel}, \citenamefont {Ludwig}, \citenamefont {Lorenz}, \citenamefont
  {Eberl},\ and\ \citenamefont {Kotthaus}}]{hüttel2005direct}%
  \BibitemOpen
  \bibfield  {author} {\bibinfo {author} {\bibfnamefont {A.}~\bibnamefont
  {H{\"u}ttel}}, \bibinfo {author} {\bibfnamefont {S.}~\bibnamefont {Ludwig}},
  \bibinfo {author} {\bibfnamefont {H.}~\bibnamefont {Lorenz}}, \bibinfo
  {author} {\bibfnamefont {K.}~\bibnamefont {Eberl}}, \ and\ \bibinfo {author}
  {\bibfnamefont {J.}~\bibnamefont {Kotthaus}},\ }\href {\doibase
  10.1103/PhysRevB.72.081310} {\bibfield  {journal} {\bibinfo  {journal}
  {Physical Review B}\ }\textbf {\bibinfo {volume} {72}},\ \bibinfo {pages}
  {081310} (\bibinfo {year} {2005})}\BibitemShut {NoStop}%
\bibitem [{\citenamefont {Kouwenhoven}\ \emph {et~al.}(1997)\citenamefont
  {Kouwenhoven}, \citenamefont {Oosterkamp}, \citenamefont {Danoesastro},
  \citenamefont {Eto}, \citenamefont {Austing}, \citenamefont {Honda},\ and\
  \citenamefont {Tarucha}}]{kouwenhoven1997excitation}%
  \BibitemOpen
  \bibfield  {author} {\bibinfo {author} {\bibfnamefont {L.}~\bibnamefont
  {Kouwenhoven}}, \bibinfo {author} {\bibfnamefont {T.}~\bibnamefont
  {Oosterkamp}}, \bibinfo {author} {\bibfnamefont {M.}~\bibnamefont
  {Danoesastro}}, \bibinfo {author} {\bibfnamefont {M.}~\bibnamefont {Eto}},
  \bibinfo {author} {\bibfnamefont {D.}~\bibnamefont {Austing}}, \bibinfo
  {author} {\bibfnamefont {T.}~\bibnamefont {Honda}}, \ and\ \bibinfo {author}
  {\bibfnamefont {S.}~\bibnamefont {Tarucha}},\ }\href {\doibase
  10.1126/science.278.5344.1788} {\bibfield  {journal} {\bibinfo  {journal}
  {Science}\ }\textbf {\bibinfo {volume} {278}},\ \bibinfo {pages} {1788}
  (\bibinfo {year} {1997})}\BibitemShut {NoStop}%
\end{thebibliography}%

\end{document}